\documentclass[preprint,12pt]{elsarticle}

\usepackage{amssymb}

\usepackage{xcolor}

\journal{Nuclear Inst. and Methods in Physics Research, A}

\usepackage{subfigure, wrapfig}
\usepackage{url}

\begin{document}

\begin{frontmatter}



\title{Characterization of  High Purity Germanium Point Contact Detectors with Low Net Impurity Concentration}


\author[lbnl,tum,mpp]{S.\ Mertens}
\author[tub,lbnl]{A.\ Hegai}
\author[ornl]{D.C. Radford}
\author[lbnl]{N.\ Abgrall}
\author[lbnl]{Y.-D. Chan}
\author[lbnl,queens]{R.D.\ Martin}
\author[lbnl]{A.W.P. Poon}
\author[tub,lbnl]{C.\ Schmitt}

\address[tub]{University of Tuebingen, Tuebingen, Germany}
\address[tum]{Technical University Munich, Department of Physics, Germany}
\address[mpp]{Max Planck Institute for Physics, Munich, Germany}
\address[lbnl]{Lawrence Berkeley National Laboratory, Nuclear Science Division, Berkeley, CA 94720, USA}
\address[ornl]{Oak Ridge National Laboratory, Oak Ridge, TN 37830, USA}
\address[queens]{Queens University, Department of Physics, Engineering Physics \& Astronomy, Kingston, ON K7L 3N6, Canada}

\begin{abstract}
High Purity germanium point-contact detectors have low energy thresholds and excellent energy resolution over a wide energy range, and are thus widely used in nuclear and particle physics. In rare event searches, such as neutrinoless double beta decay, the point-contact geometry is of particular importance since it allows for pulse-shape discrimination, and therefore for a significant background reduction. In this paper we investigate the pulse-shape discrimination performance of ultra-high purity germanium point contact detectors. It is demonstrated that a minimal net impurity concentration is required to meet the pulse-shape performance requirements.
\end{abstract}




\end{frontmatter}


\section{Introduction}
\label{sec:intro}
The main advantage of the p-type point-contact (PPC) geometry (Fig.~\ref{fig:PPC}), is to achieve a low capacitance even for large detector dimensions~\cite{Luke, Spieler}. A low capacitance allows for low noise levels, which leads to superior energy resolution and low energy thresholds. This feature makes the technology attractive for experiments in nuclear and particle physics, such as searches for dark matter, coherent neutrino-nucleus scattering, and neutrinoless double beta decay~\cite{Giovanetti:2014fhx, Abgrall:2013rze, Ackermann:2012xja, Agostini:2013mzu, Aalseth:2012if, Akimoveaao0990}. 

Beyond their superb energy resolution, PPC detectors allow for an excellent discrimination between events that deposit their energy at multiple sites in the detector volume (multi-site events) and those that deposit their energy at a single location (single-site events). This feature is critical for neutrinoless double beta decay searches in $^{76}$Ge, since it provides a means to greatly reduce events from dominant background sources, which largely manifest as multi-site events. 

The currently leading ${}^{76}$Ge-based neutrinoless double beta decay experiments, the \textsc{Majorana Demonstrator}~\cite{Abgrall:2013rze} and \textsc{Gerda}~\cite{Ackermann:2012xja} are exploiting this technology. The \textsc{Majorana Demonstrator} is using PPC detectors produced by AMETEK / ORTEC~\cite{ortec}, and \textsc{Gerda} is using a very similar type of detector, called Broad Energy Germanium (BeGe) detector produced by CANBERRA~\cite{canberra_olen,canberra_usa}.

Typically, for rare event searches, a small surface to volume ratio is advantageous. The use of large detectors minimizes surface-related background and the quantity of cables and electronics channels needed. In order to fully deplete large detectors
(with a typical mass of $\approx 800$~g)
at voltages below about 5~kV, low net active impurity concentrations of the order of $10^{10}$~cm$^{-3}$ are necessary. In this work, we show that too low a net active impurity concentration, of the order of $1\cdot10^{9}$~cm$^{-3}$, will degrade the pulse-shape discrimination performance.

Uniformity of the pulse shapes originating from events throughout the detector volume is the key requirement for a reliable pulse-shape discrimination. This feature is guaranteed by a largely uniform electric field, that strongly increases only in the close vicinity of the point contact. Too low a net impurity concentration leads to small electric fields in the corners of the detector, and therefore to an excessive variation of the field inside the detector volume. Consequently, ultra-low net impurity concentrations cause pulse shapes to be strongly dependent on the position of the interaction point in the detector volume. 

We investigate this behavior with two natural Ge PPC detectors developed by Los Alamos National Laboratory (LANL) in collaboration with AMETEK / ORTEC~\cite{ortec, XU2015807}. One of them has an average net impurity concentration of $\approx 2.5\cdot10^9$~cm$^{-3}$, whereas the other one exhibits extremely low net impurity concentration of $\approx 6\cdot10^8$~cm$^{-3}$ (between $2\cdot10^9$ and zero). The latter has significantly degraded pulse-shape discrimination performance. With dedicated measurements we demonstrate a clear event-position dependence of the pulse shapes for this detector. 

A detailed simulation, including axially symmetric field calculation and three-dimensional electron-hole propagation, is used to model this behavior. Here, we demonstrate the importance of taking into account the finite charge-cloud size and the evolution of charge-cloud shape, including the effects of self repulsion and diffusion, in order to reproduce the observed detector response.
The agreement between our model and the data confirms the impact of ultra-low net impurity concentration on pulse shapes.

This paper is structured as follows: We first describe the working principle of PPC detectors in section~\ref{sec:PPC}. In section~\ref{sec:Measurement} we give a detailed overview of the performed measurement, followed by a description of the simulation code, and a comparison of the measurements and simulations in sections~\ref{sec:Model} and \ref{sec:Comparison}. 

\section{Working Principle of PPC detectors}
\label{sec:PPC}
\begin{figure}
  \centering
  \begin{minipage}{0.6\textwidth}
   \includegraphics[width = \textwidth]{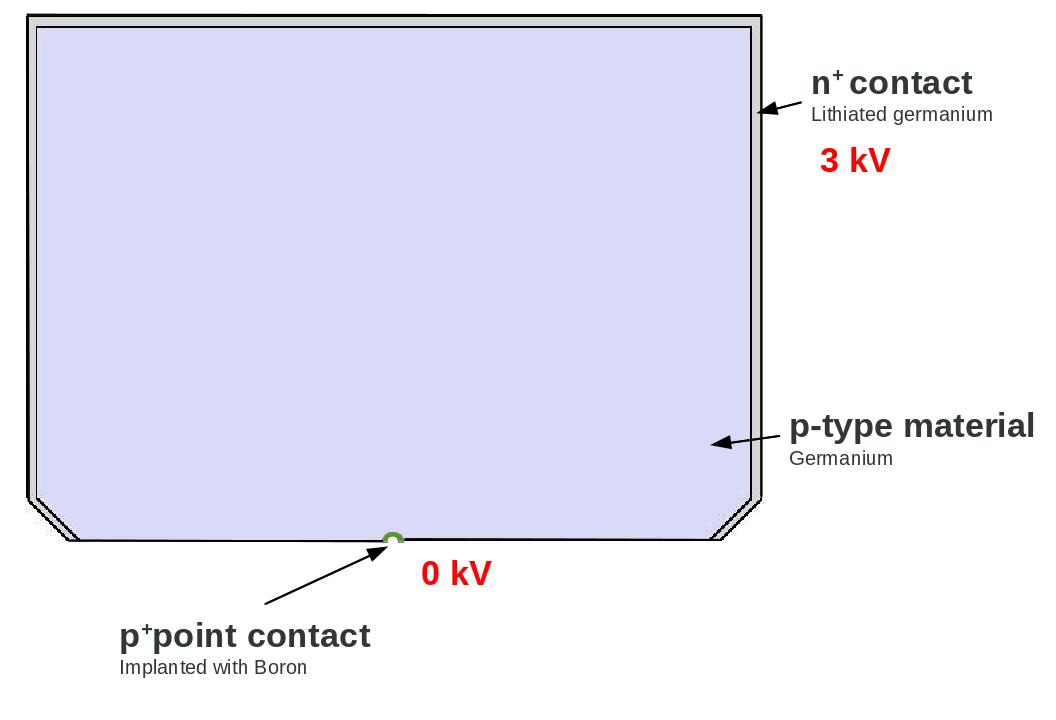}
  \end{minipage}
  \hfill
  \begin{minipage}{0.39\textwidth}
    \caption{Schematic view of a PPC Ge detector. The gray area indicates the lithiated $\mathrm{n}^+$-type contact, which is held at positive bias voltage. Here 3~kV is chosen as a typical example. The small point contact, illustrated as an arc at the bottom of the detector, is kept at zero potential. In this configuration the holes created by energy depositions in the bulk of the detector move to the point contact, whereas the electrons move to the $\mathrm{n}^+$ contact.}
    \label{fig:PPC}
  \end{minipage}
\end{figure}

The bulk of the detector is p-type germanium. Since hole trapping is generally less pronounced than electron trapping, the use of p-type material, with the collection of holes to the point contact, is advantageous for efficient charge collection, which leads to better energy resolution. Furthermore, the robust lithium $n^{+}$ contact allows for ease of handling, and prevents surface alphas from reaching the active detector volume. It typically has a thickness of around 0.5 - 1.0 mm. The small $\mathrm{p}^+$ point contact is usually created via boron implantation. The $\mathrm{n}^+$ contact is held at positive bias voltage, whereas the point contact is kept at zero potential. The surface between the point contact and the $\mathrm{n}^+$ contact is covered with a passivation layer (Fig.~\ref{fig:PPC}).

The total electric potential is the superposition of the potential created by the bias voltage and the potential arising from the net charge density of electrically active impurities. In particular, the field in the top corners (far away from the point contact) is governed solely by the net impurity concentration. The small size of the point-contact readout electrode results in sharply peaked current signals whenever holes created by an energy deposition are collected at the contact. As described in section~\ref{sec:Model} below, this is the result of two effects, namely the localized ``weighting potential'' and the strong electric field close to the contact.

\begin{figure}
  \centering
  \subfigure[]{\includegraphics[width = 0.49\textwidth]{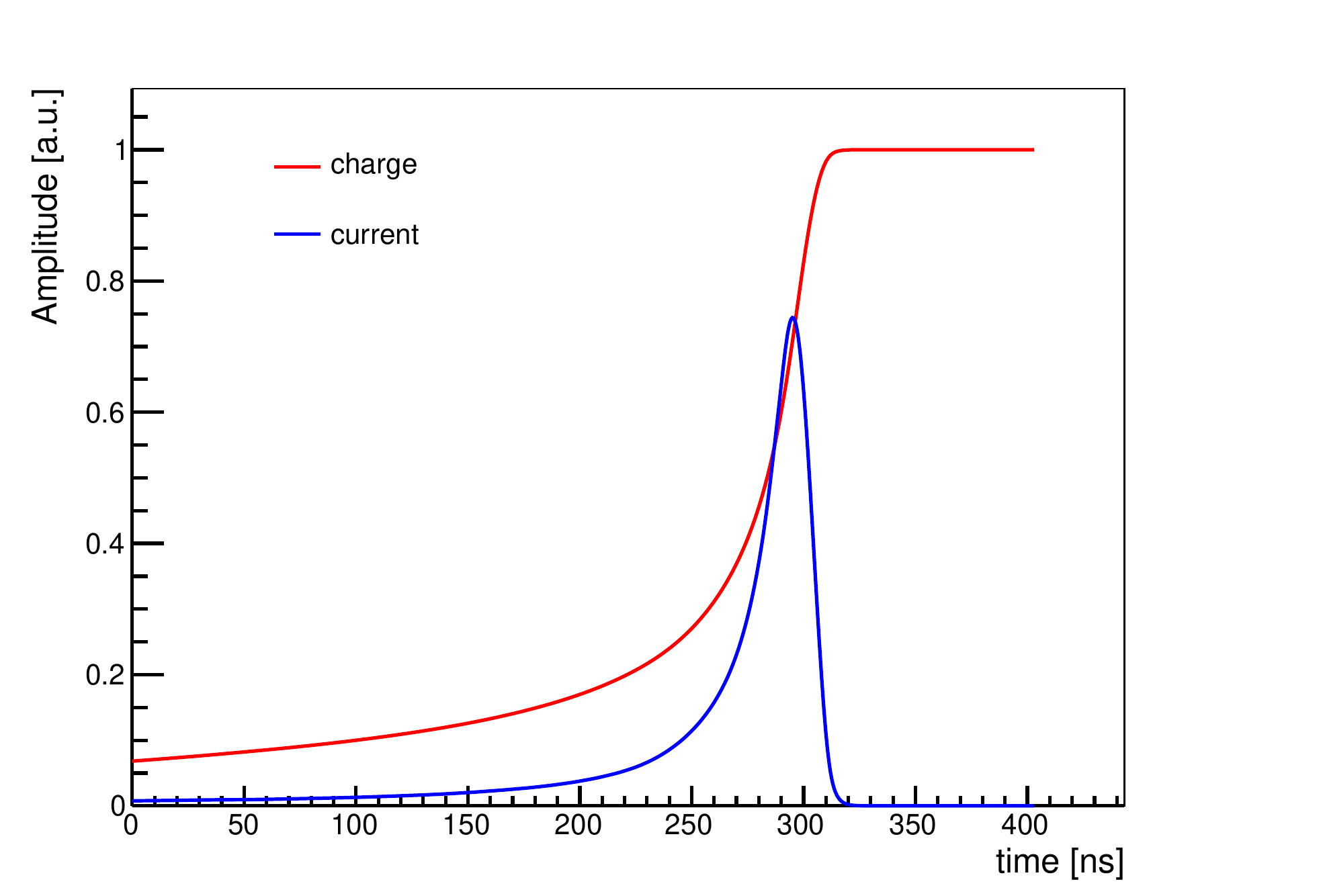}}
  \subfigure[]{\includegraphics[width = 0.49\textwidth]{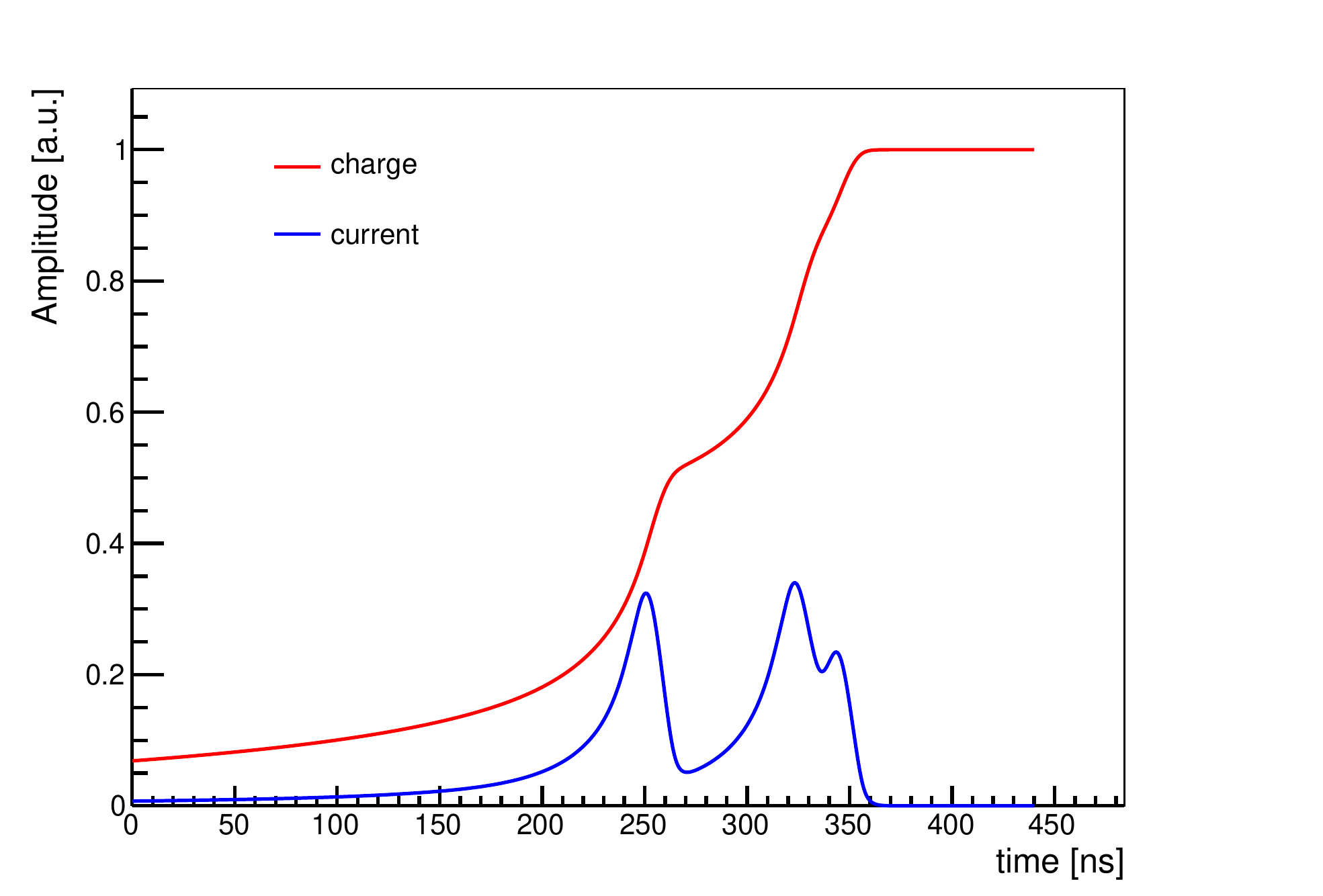}}
  \caption{(a): Simulated charge and current pulses for a single-site event. (b): Simulated charge and current pulses for a multiple-site event. The red solid line shows the charge as as a function of time, the blue line corresponds to its derivative, i.e.\ the current. Most of the signal is created only while the charge is in close vicinity to the point contact. Consequently, the ratio of maximum current to the amplitude of the signal (energy) is very similar for all single-site events and different for multiple-site events.}
 \label{fig:SSEMSE}
\end{figure}

This configuration leads to pulse shapes which are almost independent of their point of origin, but deviate from this shape when an event involves energy depositions at multiple positions in the detector (Fig.~\ref{fig:SSEMSE}). A simple discriminative property of the pulse shape is the ratio of the maximum current (A) and the amplitude (E) of the pulse. 

\section{Measurements}
\label{sec:Measurement}
Two natural Ge PPC detectors - PONaMa-I and PONaMa-II (\textbf{P}PC from \textbf{O}RTEC made from \textbf{Na}tural \textbf{Ma}terial) - were used for the studies described in this work. In these measurements, performed at Lawrence Berkeley National Laboratory, the detectors were mounted in vendor-supplied cryostats, equipped with a first amplification stage. The amplified signal was then digitized by a Struck SIS3302 digitizer card~\cite{Struck}.

In this section, we describe the general properties of the detectors, their pulse-shape discrimination performance, and the measurements that were performed in order to investigate the interaction-position dependence of the pulse shapes.

	\subsection{General detector properties}
The detectors were fabricated by the manufacturer AMETEK / ORTEC$^{\circledR}$, in Oak Ridge, TN, USA~\cite{ortec} as part of a pre-production QA run of the \textsc{Majorana Demonstrator} enriched Ge detectors. The two detectors originate from adjacent segments of a single pulled crystal, where PONaMa-I was fabricated from a part closer to the seed, and PONaMa-II from a part closer to the tail. 

In the process of crystal pulling the part close to the seed typically exhibits larger net p-type impurity concentration, as the solubility for p-type impurities is generally higher in the solid phase. The liquid phase, in contrast, has a higher solubility for n-type impurities. Hence the net p-type impurity concentration of the crystal closer to the tail, which remains in liquid phase for longer times, is typically lower. 

We estimate the net impurity concentration profile in the detector on values supplied by the detector manufacturer, scaled such that the calculated depletion voltage reproduces the measured one. The values provided by the manufacturer are based on Hall sensor measurements performed on test slices adjacent to the top and the bottom of the detector, and interpolated according to a model of the principal impurities and their segmentation characteristics. This method is prone to uncertainties of at least 20\%. In section~\ref{ssec:Model} a net impurity concentration model, used in the simulations, will be described in detail. For PONaMa-I we find a mean net impurity concentration $\rho_{\mathrm{P1}} \approx 2.5\cdot10^9$~cm$^{-3}$, and for PONaMa-II a much lower value of $\rho_{\mathrm{P2}} \approx 6\cdot10^8$~cm$^{-3}$. 

The depletion and operating voltages of PONaMa-I (PONaMa-II) are low at V$_{\mathrm{dep}}$ = 880 V and V$_{\mathrm{op}}$ = 2000 V (V$_{\mathrm{dep}}$ = 450 V and V$_{\mathrm{op}}$ = 3000 V), respectively. Both detectors showed very low capacitance of less than 2~pF and low leakage currents of less than 7~pA at depletion. An excellent energy resolution over a large energy range was found for both detectors. Table~\ref{tab:properties} summarizes their characteristics and general performance. 

\begin{table}
\begin{center}
\caption{Characteristics and general performance parameters of PONaMa-I and II. The energy resolution is evaluated as the full width half maximum (FWHM) at the 1.33~MeV gamma line of $^{60}$Co and the 59.9~keV line of $^{241}$Am. Additionally, the width of the pulser signal introduced from an external pulser via the high-voltage line of the detector is given.}
\begin{tabular*}{\textwidth}{@{\extracolsep{\fill}}p{7cm}rr}
\hline
  & PONaMa-I	&  PONaMa-II \\ 
\hline \hline
Height & 50.5~mm & 47~mm  \\
Radius & 34.5~mm & 34.5~mm\\
Mass & 1~kg & 0.93~kg \\
Leakage current  & 3.4 pA & 6.4 pA\\
Capacitance  &  1.64(8) pF        &   1.56(8) pF  \\
Operating voltage   & 2000 V & 3000 V  \\
Depletion voltage   & 880 V & 450 V \\
FWHM @ 1.33 MeV   & 2100 eV & 2000 eV \\
FWHM @ 59.5 keV    & 600 eV  & 670 eV\\
Pulser FWHM             & 490  eV  & 540 eV\\
\hline
\end{tabular*}
\label{tab:properties}
\end{center}
\end{table}

	\subsection{Pulse-shape performance}

To determine the effectiveness of single-site and multi-site event discrimination, a $^{232}$Th calibration source was used. In the decay of $^{208}$Tl a high energy (2.615~MeV) gamma ray is produced, which has a high probability of producing an $e^{+}$$e^{-}$ pair in the Ge detector. The subsequent two 511-keV gammas produced by annihilation of the positron can either fully deposit their energy in the detector (full energy peak), only one gamma deposits energy in the detector (single-escape peak), or both gammas leave the detector without interaction (double-escape peak). The latter generates almost exclusively single-site events, whereas the single-escape peak is populated with multi-site events. These two event classes are ideal to evaluate the efficiency of pulse-shape discrimination. 

\begin{figure}
  \centering
  \includegraphics[width = 0.6\textwidth]{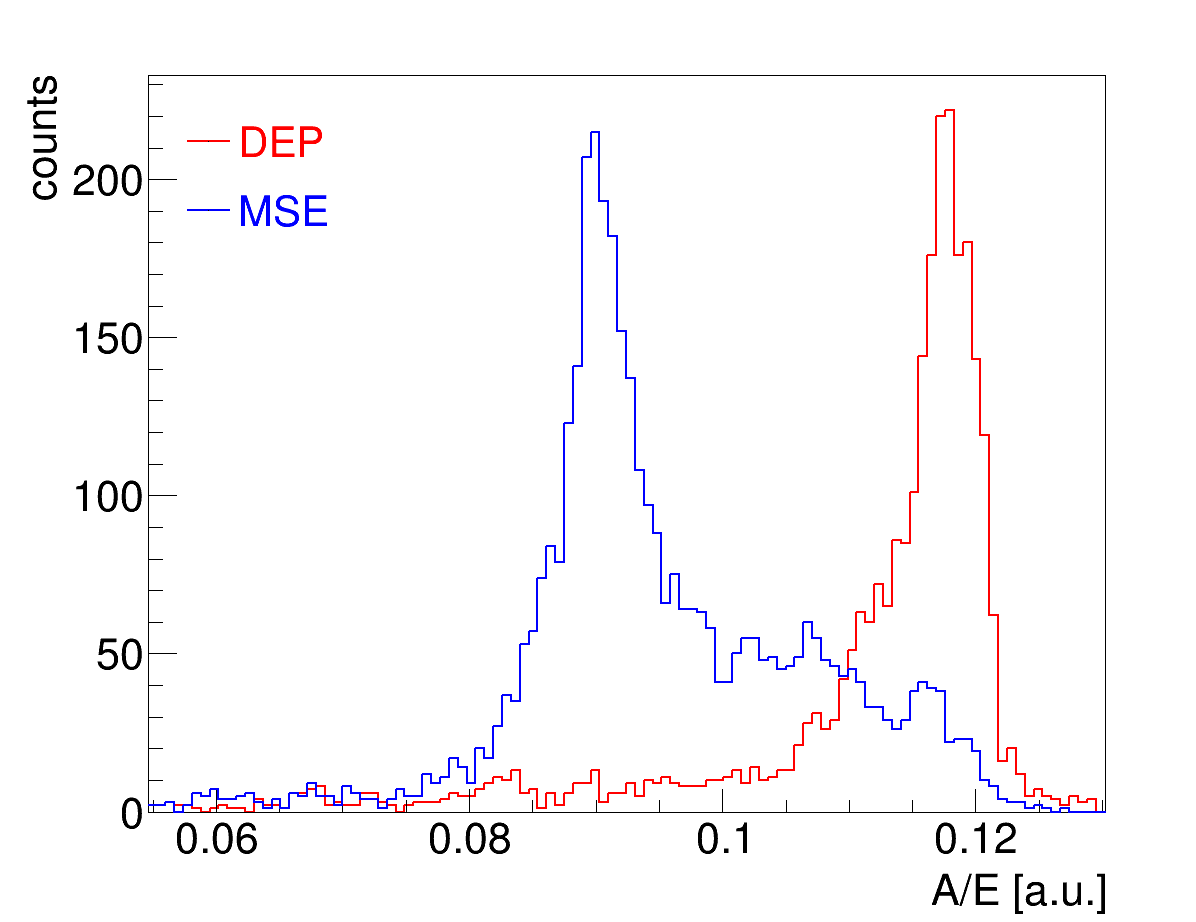}
  \caption{Measured A/E distribution for double-escape peak (DEP) events in red and multi-site events (MSE), based on the single-escape peak in blue. The DEP distribution is dominated by single-site events since the energy is deposited by the electron and the positron in a small volume. The single-escape peak on the other hand is multi-site dominated since additional energy is deposited by a gamma.}
 \label{fig:AoverEDistrib}
\end{figure}

As described above, a simple discriminating property of the pulse shape is the ratio of the maximum current (A), i.e. the maximum of current arriving at the point contact per unit time, and the total amplitude (E) of the pulse.  The A/E distribution of single-site events (SSE) is narrow and almost independent of energy, whereas the A/E distribution of multi-site events (MSE) is a broad and centered at a value smaller than that for SSE (Fig.~\ref{fig:AoverEDistrib}). The discrimination between SSE and MSE is performed by a slightly energy-dependent cut on the A/E value. A detailed description of this procedure can be found in~\cite{Agostini:2013jta}.

By fixing the A/E cut parameter value to set the acceptance of the double-escape events at $90\%$, and then measuring the acceptance at the single-escape peak, the cut efficiency $\epsilon$ for the multi-site events is determined. Natural limitations of this approach include the fact that the energy gates on the DEP and SEP peaks include some underlying continuum from Compton scattering of the 2614.5-keV gamma rays. In addition, some events collected in the double-escape peak can in fact be multi-site; forward-angle Compton scattering of the 2614.5-keV gamma, followed by pair production at a different site, is unlikely but not excluded. For the single-escape peak, some multi-site events, where charge carriers from different interaction points accidentally have almost identical drift times to the point contact, can also occur. However, the pulse-shape analysis (PSA) cut is generally very selective; the PPC detectors in the {\sc{Majorana Demonstrator}} array typically have $\epsilon$ values in the range $6\%$ to $10\%$ \cite{Majorana_PPC}.

In our test, PONaMa-I showed an excellent multi-site event rejection efficiency of $\epsilon=9.6\pm0.9\%$, whereas PONaMa-II only reached $\epsilon=35.2\pm0.2\%$. This result led to a research effort aimed at understanding its causes and ensuring that such poor performance could be avoided in detectors used for the \textsc{Majorana Demonstrator} experiment.

The fact that the pulse-shape discrimination performance is degraded while the energy resolution is unaffected, indicates that the charge is fully collected, but with a different profile of collection times. The collection time is defined by the time to collect 30\% to 80\% of the charge, and is not to be confused with the drift time of the charges.  As described above, low electric fields in the corners of the detector, which can be caused by extremely low net impurity concentrations, can lead to growth of the charge-cloud size~\footnote{A certain number of electron-hole-pairs is created in each energy deposition in the detector. The volume in which these charge carriers are created is referred to as charge cloud size.} and an increased collection time of charges created in these regions. The dependence of collection time on the position of energy deposition can widen the A/E distribution, and cause the observed degraded discrimination performance. Specific measurements to test this hypothesis are described in the following. 
	
	\subsection{Position dependence of pulse shapes}
Two specific measurements were performed on both detectors in order to test whether the A/E parameter depends on the position of energy deposition. The first was a radial scan of the top surface of the detector and the second a drift-time measurement. 

\paragraph{Radial scan}
To study the position dependence of the A/E parameter, a collimated $^{241}$Am source was placed on the top surface of the pop-top cryostat at 2~cm distance to the detector top surface. The beam spot, which was less than 2mm in diameter, was moved in 1~mm steps from the outer cryostat radius at X~=~0~cm, across the center of the detector, to the opposite outer cryostat radius at X$~\approx65$~mm. The mean free path in Ge of 59.9~keV gammas from the $^{241}$Am decay is only about 1~mm, such that the characteristics of the pulse shapes close to the surface can be probed.

\begin{figure}
  \centering
  \subfigure[]{\includegraphics[width = 0.49\textwidth]{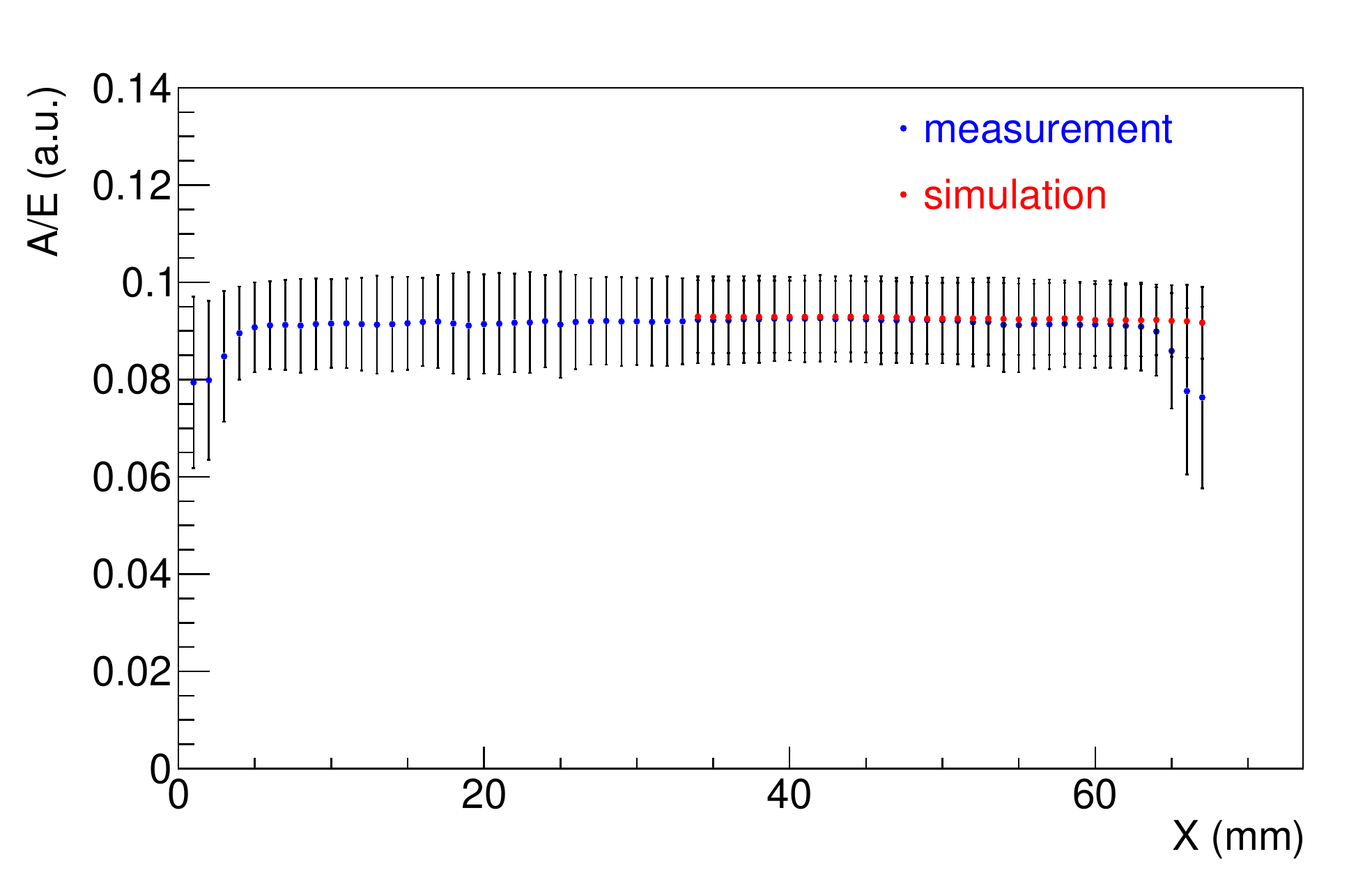}}
  \subfigure[]{\includegraphics[width = 0.49\textwidth]{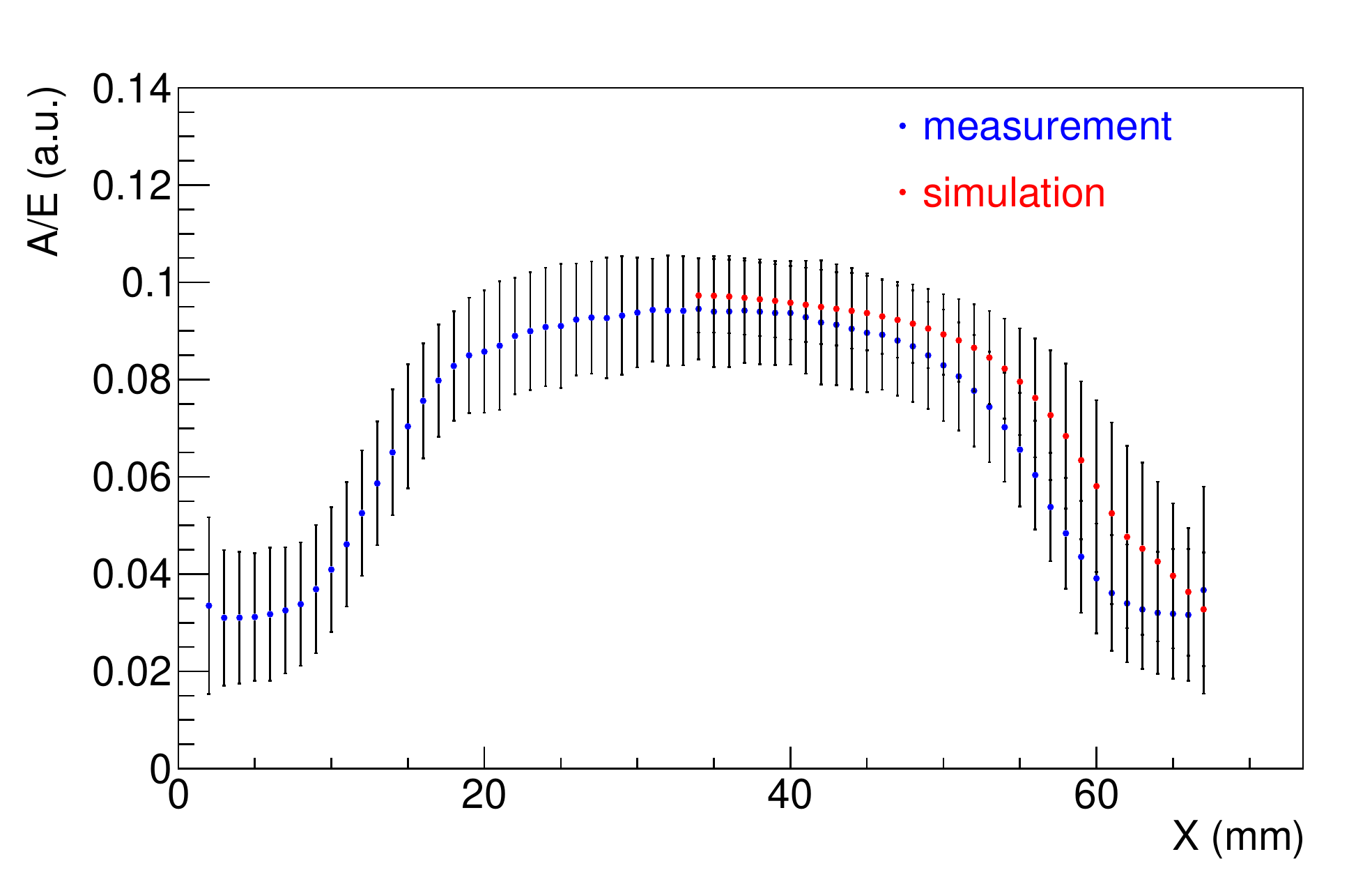}}
  \caption{Simulated (red points) and measured (blue circles) dependence of A/E on the horizontal position X of PONaMa-I (a) and PONaMa-II (b). The center of the detector is at X = 34~mm. As described in the text, PONaMa-II shows a degraded A/E value at larger radii. The good agreement between measurement and simulation (detailed in section~\ref{sec:Model} and~\ref{sec:Comparison}) demonstrates that this behavior can be explained by low net-impurity concentration and resulting low fields in the corners.}
 \label{fig:radial_scan}
\end{figure}

For PONaMa-II, which has an ultra-low net impurity concentration, we expect a less steep slope of the rising edge of the pulse shapes (i.e. a smaller value for A) for charges created in the corner (i.e. large radii) of the detector. In fact, the radial scan of PONaMa-I revealed almost no position dependence, whereas in the case of PONaMa-II, the A/E parameter dropped drastically as a function of radius (Fig.~\ref{fig:radial_scan}).

\paragraph{Drift time}
Another way to identify the position of charge creation is by measuring the drift time. A point-contact detector itself does not provide precise information on the drift time of charges, since the bulk of the signal is only created when the charges are close to the point contact~\cite{MARTIN201298}. In order to acquire this additional information, we used a collimated $^{22}$Na source and placed it between the Ge detector and a NaI(Tl) scintillator detector. $^{22}\mathrm{Na}$ emits two back-to-back 511-keV gammas from positron annihilation, which can hit the two detectors simultaneously. The trigger time of the fast NaI(Tl) scintillator was used as a proxy for the start of the drift $t_\mathrm{init}$ and the PPC detector determines the arrival time $t_\mathrm{final}$ of the charges at the point contact.

\begin{figure}
  \centering
  \subfigure[]{\includegraphics[width = 0.49\textwidth]{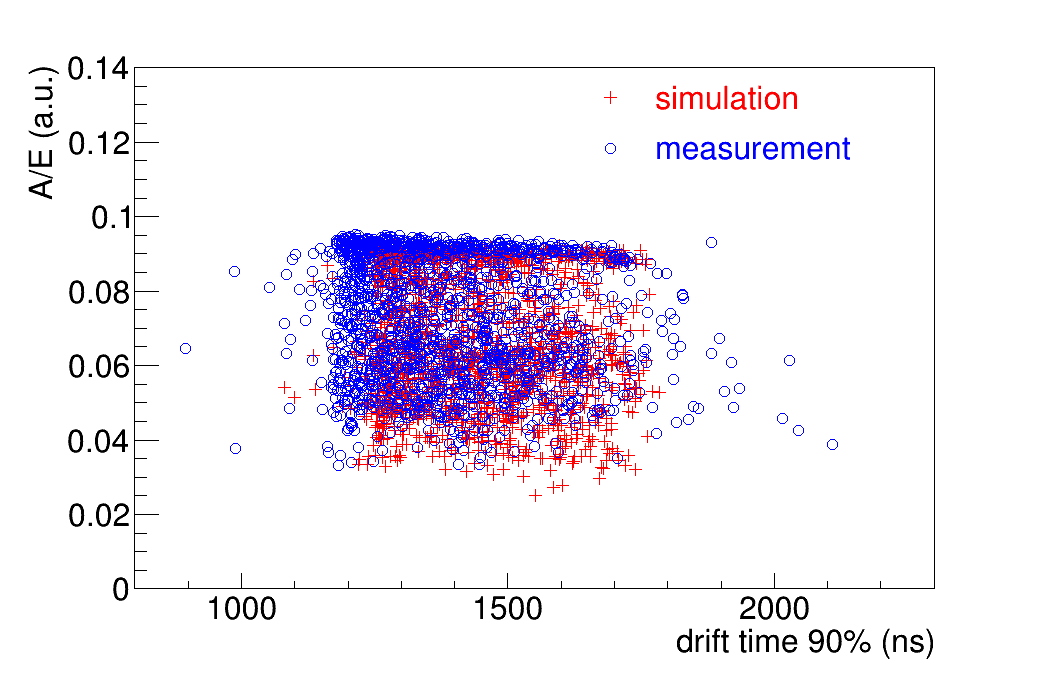}}
  \subfigure[]{\includegraphics[width = 0.49\textwidth]{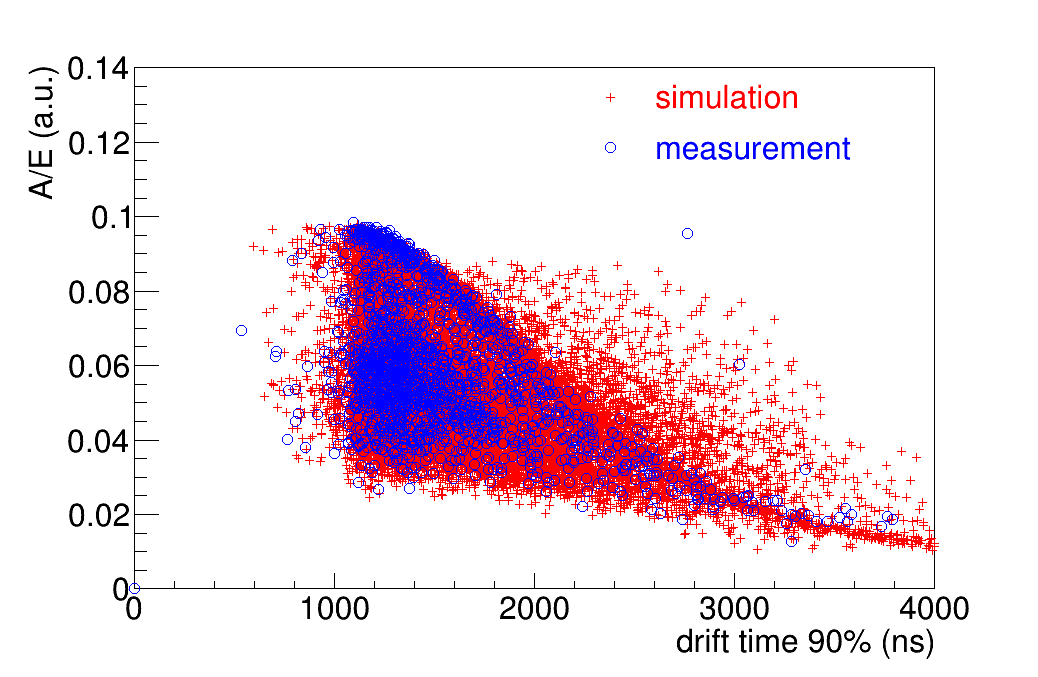}}
  \caption{Simulated (red dots) and measured (blue dots) dependence of A/E on the drift time of charges for PONaMa-I (a) and PONaMa-II (b). The drift time displayed on the x-axis is calculated as the time elapsed between the time the germanium pulse reaches 90\% of its amplitude and the time the NaI pulse reaches 50\% of its amplitude. As described in the text, the A/E value of PONaMa-II is degraded for long drift times. As in the case of the position scan (Fig.~\ref{fig:radial_scan}), the good agreement with the simulation demonstrates that the behavior can be explained by low net-impurity concentrations and resulting low fields in the corners of PONaMa-II. }
 \label{fig:coincidence}
\end{figure}

To probe the low field region, the gamma source was placed at the top-side edge of the Ge crystal. Charges created in this area have the longest drift time. In the case of low net impurity concentration, we again expect the pulse shape of charges created in the corner to have a smaller value for the parameter A. The measurement revealed that indeed the A/E value decreases as a function of drift time in the case of PONaMa-II and stays constant in the case of PONaMa-I (Fig.~\ref{fig:coincidence}).
	 
\begin{figure}
  \centering
  \subfigure[]{\includegraphics[width = 0.49\textwidth]{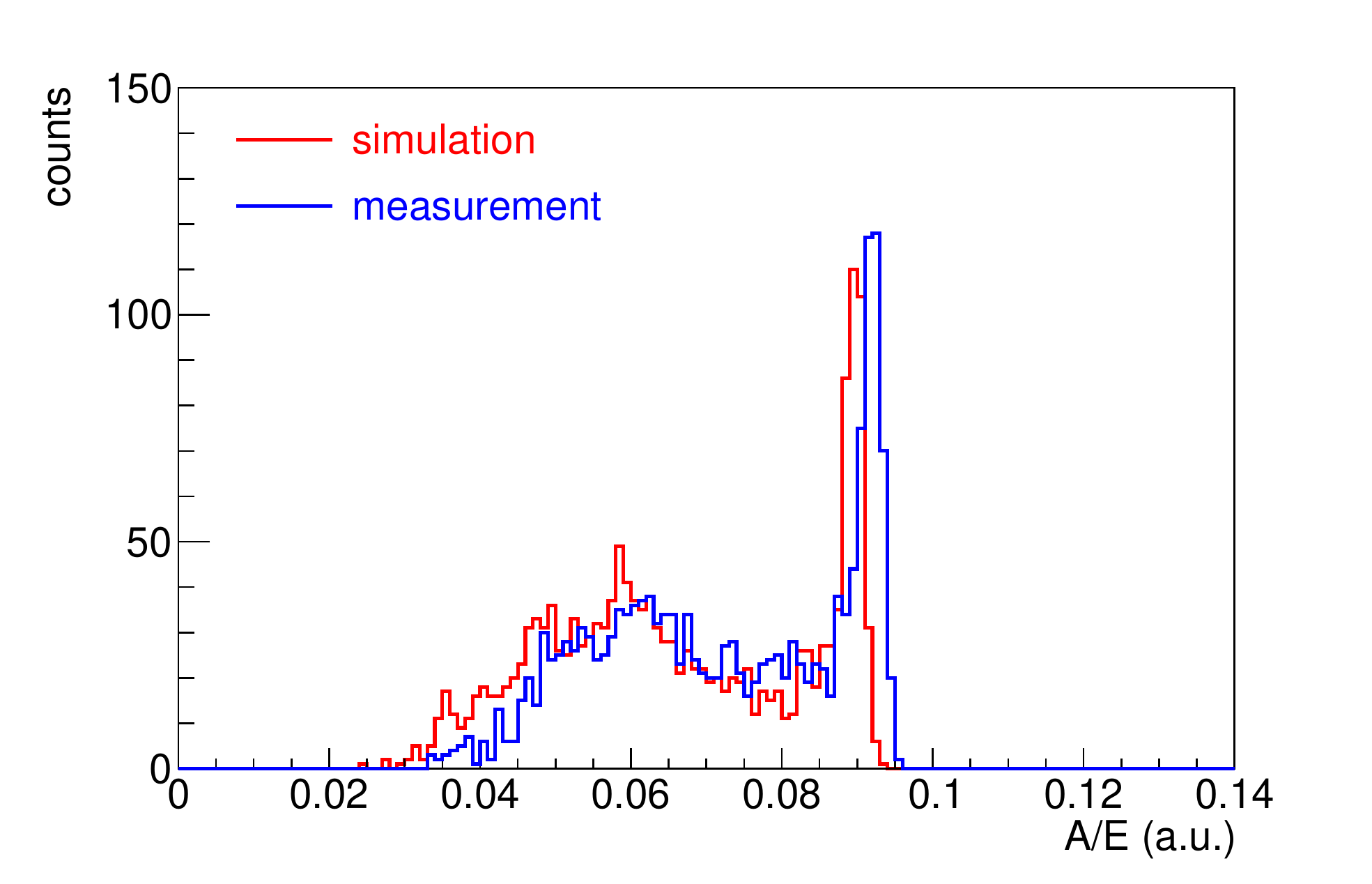}}
  \subfigure[]{\includegraphics[width = 0.49\textwidth]{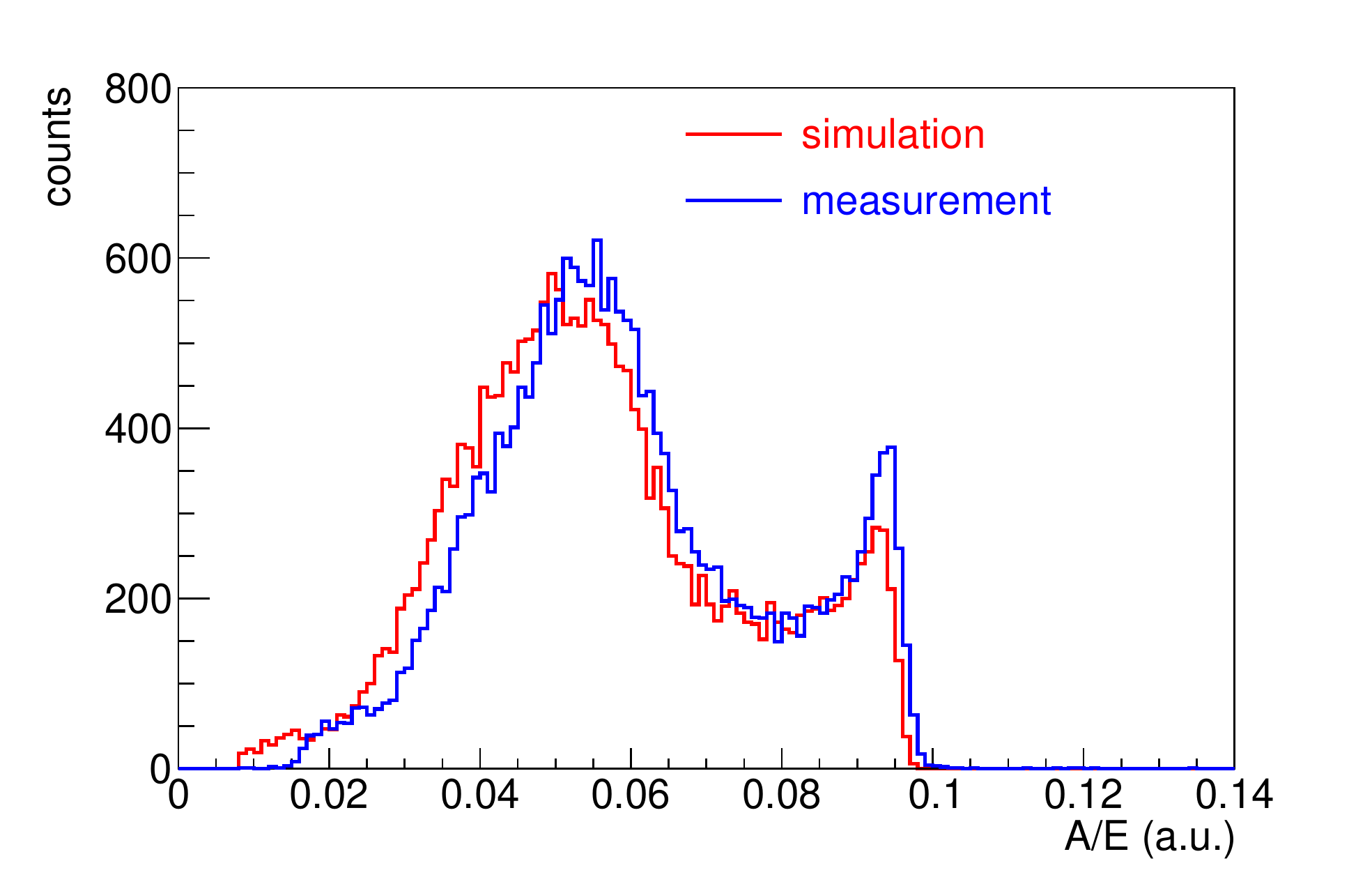}}
  \caption{Simulated (red line) and measured (blue line) A/E distribution for the measurement with the $^{22}$Na source for PONaMa-I (a) and PONaMa-II (b). A region of interest (ROI) cut at $511\pm5$~keV was applied. PONaMa-II shows significantly degraded A/E values. The drift time measurement, displayed in Fig.~\ref{fig:coincidence}, demonstrates that the low A/E values are associated to events with long drift times, and hence events that originated in the top side edge of the detector far from the point contact.}
 \label{fig:AE_sim}
\end{figure}

\section{Pulse-Shape Simulation}
\label{sec:Model}
The Ge-detector signal simulations used in this work were generated by the package {\tt{mjd\_siggen}}~\cite{radware}. This package comprises two separate parts. The first of these, {\tt{mjd\_fieldgen}}, is a stand-alone code used to calculate the electric field and ``weighting potential'' of the point contact inside the detector. These results are then used as input to the second part, the {\tt{mjd\_fieldgen}} library. There, the field maps are combined with electron- and hole-mobility information to track the drift of charges produced by interactions in the detector, and to calculate the resulting signal through the Shockley-Ramo theorem~\cite{Ramo}\cite{Shockley}\cite{He2001250}.

Both codes have been validated by comparison to exact calculations and comparison to data. In particular, {\tt{mjd\_fieldgen}} was precisely validated by algebraic calculations for coaxial and planar detectors, for both the calculation of the potentials and the capacitances. {\tt{mjd\_fieldgen}} was validated by comparisons to ADL and to measured signals for a variety of segmented coaxial and point-contact detectors.

\subsection{Electric Field and Weighting Potential}
The field generation code {\tt{mjd\_fieldgen}} calculates the electric and weighting potentials inside the detector through a numerical relaxation algorithm. For PPC detectors, the cylindrical symmetry of the detectors can be exploited to reduce the problem to two dimensions. The calculation is performed on a regular (r,z) grid. For the calculations reported here, a grid size of 0.1 mm $\times$ 0.1 mm was used.

The relaxation algorithm requires boundary conditions. For the passivated surface, the code assumes reflection symmetry about that surface. This is equivalent to the requirement that, in the absence of any surface charge on the passivation, the electric field at the passivated surface is parallel to that surface. 

In addition to the detector geometry, the code requires information on the spatial distribution of the net electrically-active impurity concentration inside the germanium. The code allows for a bulk net impurity concentration that varies quadratically with height $z$, and with a variable power in radius $r$. Furthermore, a surface charge density resulting from the passivation layer can be included. The simulation settings, used in this work, are summarized in table~\ref{tab:simulation_settings}.

When calculating the electric potential, {\tt{mjd\_fieldgen}} iteratively finds any undepleted regions of the detector volume. An undepleted voxel is indicated by a local minimum (or maximum for n-type material) in the electrical potential created by the impurities. Undepleted regions of a semiconductor detector have no net space charge and no field. Therefore, in {\tt{mjd\_fieldgen}}, the local net impurity concentration is set to zero for any voxels that are identified as undepleted, and they are assigned a single common potential. If there is a region of undepleted voxels that are connected to the readout contact (and therefore at zero electrical potential), they are treated as being part of the contact during the weighting potential calculation. This procedure allows for realistic simulations of signals even for detectors that are not fully depleted. Finally, {\tt{mjd\_fieldgen}} calculates the capacitance of the readout contact. 

  \subsection{Signal Generation}
The calculation of the signal induced by charge carriers inside the detector requires not only the weighting potential but also their position as a function of time. {\tt{mjd\_siggen}} combines the fields calculated by {\tt{mjd\_fieldgen}} with literature values for electron and hole mobilities~\cite{PhysRevB.24.1014}\cite{OMAR19871351} to calculate the charge trajectories. Both the temperature dependence and the crystal-axis dependence of the charge velocities are taken into account. 

For a given field strength, the charge velocities are anisotropic both in magnitude and direction with respect to the crystal lattice axes. Without additional constraints, however, the three scalar values of the velocity reported in the literature are insufficient to define the velocity vector for all field directions. In the {\tt{mjd\_fieldgen}} charge-drift model, which was developed by I-Yang Lee for GRETINA~\cite{IYLee}, the velocity tensor is required to be irrotational; this provides the additional constraints needed to determine the effect of the crystal orientation on the velocity. 

As the charge trajectories are computed, the induced signals are calculated for both the holes and electrons using the Shockley-Ramo theorem~\cite{He2001250}, and summed to give the total signal. This proceeds until both polarities of charges have been collected at their respective contacts. Finally, the effect of the preamplifier bandwidth is applied, modeled as a single RC-integration time constant.
  
  \subsection{Charge cloud evolution}
There are three different effects that can modify the size and shape of a charge cloud as it moves through a germanium detector: 1) Diffusion of charges, 2) charge self-repulsion, and 3) acceleration and deceleration of the charges in the detector. All are included in {\tt{mjd\_siggen}}. 

The initial shape of the charge cloud is assumed to be approximately spherical and Gaussian, with an assumed initial width. The shape of the cloud - its dimensions parallel to and transverse to its motion - is then allowed to change as it is accelerated and/or decelerated by the varying electric field. In order to simplify the calculation, and keep the computation time to a reasonable limit, the charge trajectories are computed only for the centroid of the charge cloud. Finally, once the centroid of the charge cloud arrives at the point contact, the effect of its longitudinal size is applied to the final signal by convolving in the time domain with a Gaussian of width $\delta t = S / v$, where $S$ is the final longitudinal size and $v$ is the final velocity of the cloud.

Diffusion of charges is treated by assuming a growing Gaussian width of the charge cloud of
$\sigma = \sqrt{2  D  t}$,
where $D$ is the diffusion coefficient and $t$ is time. The value of $D$ depends on the electric field strength, and different longitudinal and transverse coefficients are applied separately to the longitudinal and transverse dimensions.

Charge self-repulsion depends on both the energy and the initial size of the cloud, since a small, dense charge cloud will have stronger space-charge effects than a diffuse cloud. The rate of growth is modeled taking into account both the electric field of the crystal and the field generated by the cloud. The latter depends on both the length and width of the cloud, so a proper two-dimensional treatment is required. When the space-charge effects are very large, they tend to drive the cloud shape away from a Gaussian distribution, while diffusion tends to restore the Gaussian shape. In {\tt{mjd\_siggen}}, the non-Gaussian shape is ignored; this is a very good approximation for all but very short-range, high-energy depositions, such as those created by alpha particles.

Finally, the effects of acceleration and deceleration on the longitudinal size are included. As the cloud accelerates, it spreads out in such a way that the trailing charges always follow the leading charges by a fixed time interval (neglecting the above effects of diffusion and repulsion). That is, $S/v$ is constant, where again $S$ and $v$ are the longitudinal size and speed, respectively. It is evident that this effect can produce extremely elongated charge clouds for events where the energy deposition occurs in a low-field region, and the initial value of $v$ is therefore very small.

As the trajectory of the charge centroid is computed, the contributions of each of these three effects to the longitudinal and transverse sizes are evaluated at each time step. The effects are added in quadrature to obtain the total width of the charge cloud. Only the final longitudinal size and velocity of the charges arriving at the point contact are used to determine the time-domain convolution function for the final signal.

\section{Comparison of model to data}
\label{sec:Comparison}
In this section, we compare the measurement results described in section~\ref{sec:Measurement} to simulation. The model is composed of two parts, first the simulation of charge deposition in the Ge crystal obtained by the GEANT-based~\cite{Agostinelli2003250, Allison2016186} simulation software MaGe~\cite{MaGe}, and second the pulse-shape simulation with {\tt{mjd\_siggen}} described in the previous section~\ref{sec:Model}.

	\subsection{Simulation settings}
	\label{ssec:Model}
\begin{figure}
  \centering
  \includegraphics[width = 0.6\textwidth]{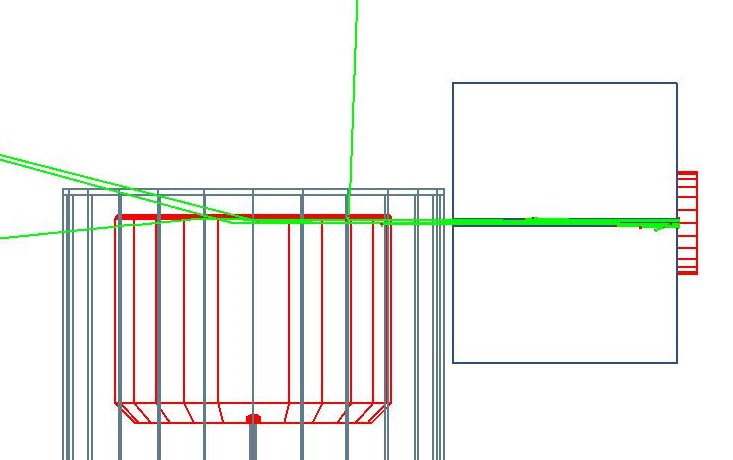}
  \caption{Visualization in MaGe of the collimated $^{22}$Na simulation. The black and red shapes display the cryostat with the detector, the $^{22}$Na button source, and the lead collimator. The green lines indicate gammas that interact in the Ge detector, mostly in the top right corner.}
 \label{fig:MaGe_side_edge}
\end{figure}

	 Geant-4 was used to simulate the locations and the energy depositions of gamma-ray interactions in the Ge crystal. The Ge detector geometry is defined via MaGe, which provides a convenient interface to Geant-4. In accordance with the measurements, we performed MaGe simulations of a $^{232}$Th source, a collimated $^{241}$Am source, and a collimated $^{22}$Na source to obtain the positions and energy depositions in the Ge crystal. Figure~\ref{fig:MaGe_side_edge} displays the simulated geometry. All energy depositions within a 1-mm sphere are combined to one interaction point and then used in the pulse shape simulation.   

	 Most importantly for this work, a detailed simulation of the signal created by each energy deposition in the crystal volume is performed. As described in section~\ref{sec:Model} a finite charge-cloud size and the evolution of the charge cloud within the Ge crystal are taken into account. 
     
\begin{figure}
  \centering
  \subfigure[]{\includegraphics[width = 0.3\textwidth]{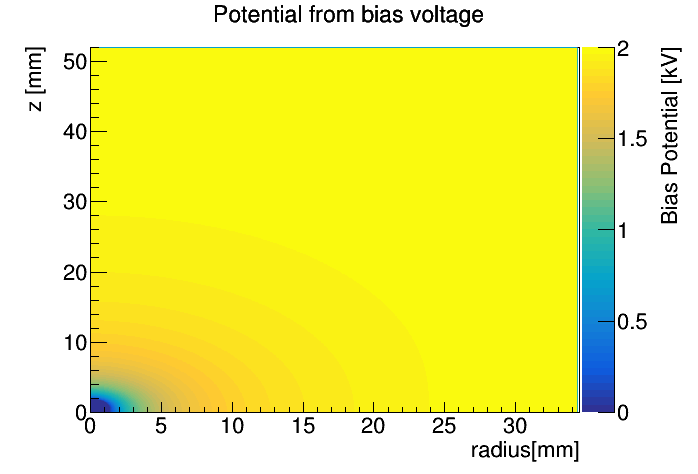}}
  \subfigure[]{\includegraphics[width = 0.3\textwidth]{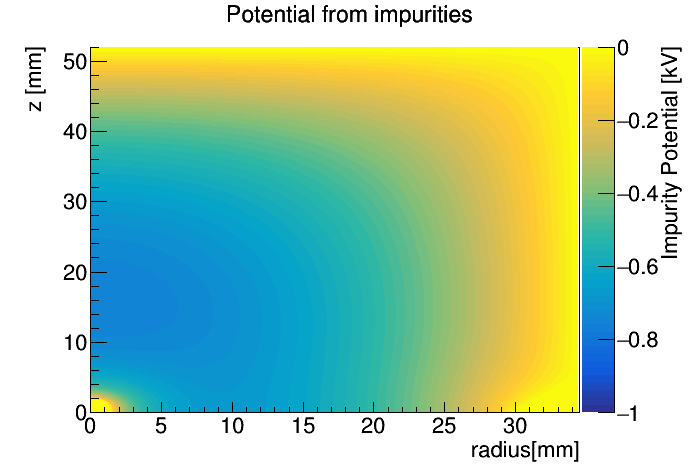}}
  \subfigure[]{\includegraphics[width = 0.3\textwidth]{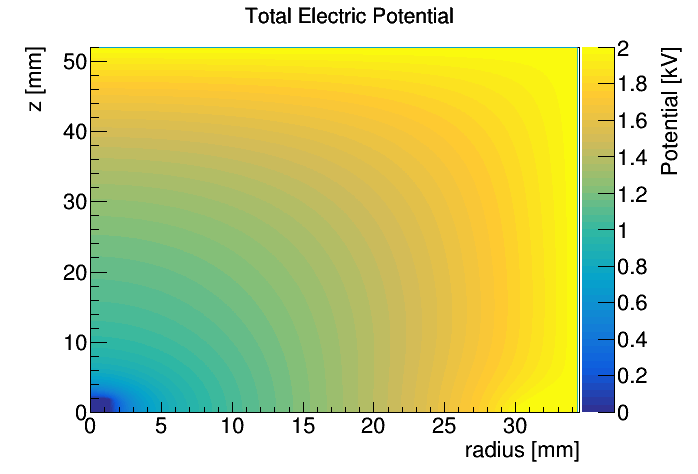}}
  \subfigure[]{\includegraphics[width = 0.3\textwidth]{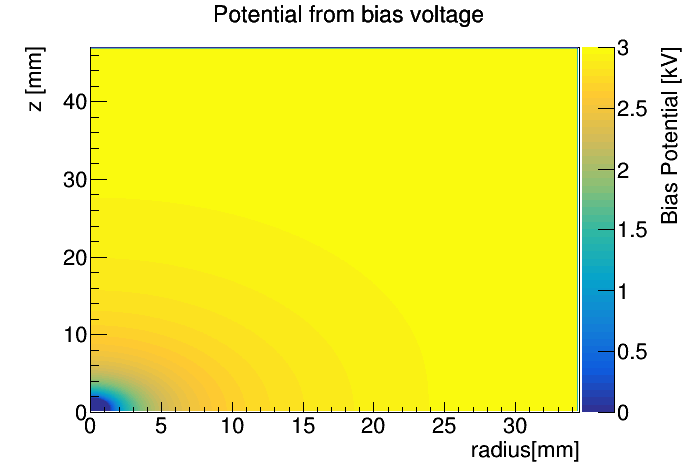}}
  \subfigure[]{\includegraphics[width = 0.3\textwidth]{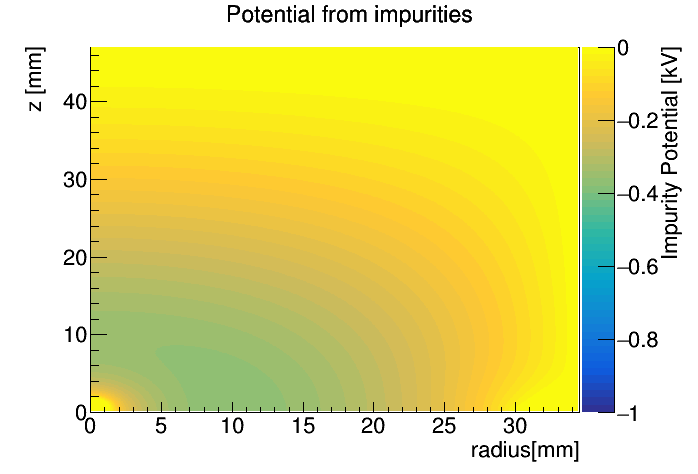}}
  \subfigure[]{\includegraphics[width = 0.3\textwidth]{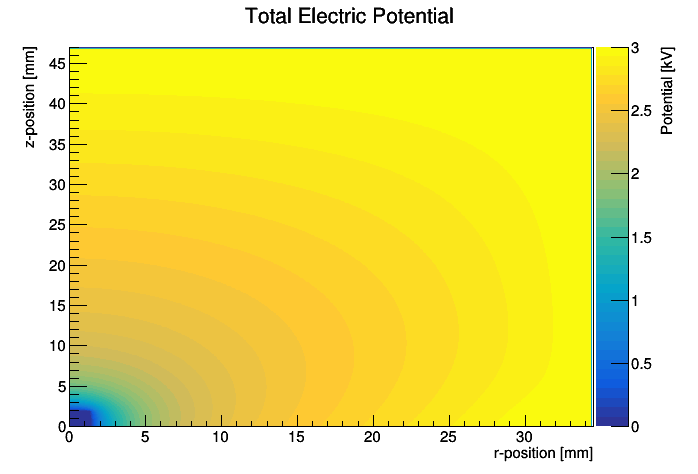}}
  \caption{Simulated electric potential in PONaMa-I. (a) Potential generated by bias voltage, (b) Potential generated by net impurity concentration, (c) superposition of (a) and (b). Simulated electric potential in PONaMa-II. (d) Potential generated by bias voltage, (e) Potential generated by net impurity concentration, (f) superposition of (d) and (e).}
 \label{fig:Field_P}
\end{figure}

     The input parameters describing the detector (detector geometry, operating voltage, etc.) were chosen according to table~\ref{tab:properties}. Figure~\ref{fig:Field_P} shows the simulated electric potential in both detectors. In this simulation no surface charge on the passivated layer was included.  
     
     The parameters describing the impurity profile of detector are given in table~\ref{tab:simulation_settings}. As mentioned above, we base our assumption on the mean net impurity concentration on the depletion voltage of the detector. To obtain the distribution of the concentration within the detector we rely on the values provided by the manufacturer. For simplicity, we assume a linear change of the concentration from the bottom to the top of the detector. By doing so, we find in case of PONaMa-II, a transition from p-type to n-type close to the top. Additionally, we allow for a slight radial dependence, which however, has a minor effect on the results.
     
     To simulate the signal amplification a preamplifier integration time (26 ns for PoNaMa-I and 21 ns for PoNaMa-II) is assumed. The simulated pulse-shapes are converted to the same data format as the measured data and are analyzed with the same analysis software. During this process white noise, based on the measured noise level, was added to the pulse-shapes to make the results more comparable.
	 
    \begin{table}
\begin{center}
\caption{Relevant simulation settings for {\tt{mjd\_fieldgen}} and {\tt{mjd\_siggen}}. {\tt{mjd\_fieldgen}} assumes a radial and longitudinal dependence of the net-impurity concentration of the form $\rho(r,z) = - ( \rho_0 + \rho_1 \times z + \rho_2 \times z^2 + \rho_r \cdot (\frac{r}{R})^{n} )$, where $R$ is the detector radius. The minus sign in $\rho(r,z)$ arises from the fact that the crystal is p-type.}
\begin{tabular*}{\textwidth}{@{\extracolsep{\fill}}p{7cm}rr}
\hline
  & PONaMa-I	&  PONaMa-II \\ 
\hline \hline\\
$\rho_0$   & $2.92\cdot10^9 ~\mathrm{cm}^{-3}$ & $ 2.4\cdot10^9 ~\mathrm{cm}^{-3}$  \\
$\rho_1$   & $- 0.9\cdot10^8$~cm$^{-4}$ & $- 7.4\cdot10^8$~cm$^{-4}$ \\
$\rho_2$   & $ 0 $ & $ 0 $ \\
$\rho_r$   & $1.0\cdot10^9$ & $-7.5\cdot10^8$ \\
$n$  & 6 & 4 \\
\hline
\end{tabular*}
\label{tab:simulation_settings}
\end{center}
\end{table}

	\subsection{Simulation results}
    
As a first step, the general pulse-shape discrimination performance was simulated. To this end the following daughter elements of ${}^{232}$Th: ${}^{212}$Bi, ${}^{208}$Tl, and ${}^{228}$Ac were implemented to emulate the realistic case in which the ${}^{228}$Ac line overlaps with the relevant double-escape peak of ${}^{208}$Tl. According to the experimental setup, the ${}^{232}$Th source was simulated at a distance of 10~cm from the detector. Here, an initial charge-cloud size of 0.2~mm was chosen. In agreement with the measurement, the simulations reveal a significantly degraded performance for PONaMa-II and compared to PONaMa-I. Table~\ref{tab:lines} summarized the results.

Secondly, the radial scan was simulated. Here, a collimated $^{241}$Am source at 5~cm distance (according to the length of the collimator) to the detector was implemented, and only the relevant gamma line at 59.5~keV was considered. For this gamma energy an initial charge-cloud size of 0.01~mm was assumed. Figure~\ref{fig:radial_scan} shows a very good agreement between simulation and measurement. Both detectors show a stable count rate in the 59.5 ~keV ROI which indicates a homogeneous charge collection efficiency. But as in the measurement, for PONaMa-II the A/E value drops drastically as a function of radius of the detector, whereas for PONaMa-I it stays rather constant. The net impurity distribution for PONaMa-II at the top of the detector is not completely known which results in a non-perfect model of the field in this region. This is the main reason for the deviation between simulated and measured A/E values at the outermost radii. 

Finally, the drift time measurement was reproduced via simulation. The relevant 511~keV line of a collimated ${}^{22}$Na source was simulated at a distance of 7.8~cm from the detector (according to the length of the collimator). Here, we assume an initial charge cloud size of 0.2~mm. As the drift time is known in the Ge signal simulation, a simulation of the NaI detector is not necessary. Due to time delays in the electronics of the NaI detector, the simulated and measured drift times can deviate by a constant factor. Since, for our study, only the relative dependence of the A/E parameter on the drift time is relevant, we allow in our analysis a shift of the average drift time to match that of the data. As in the measured data, we observe in the simulation a strong dependence of the A/E parameter on the drift time for PONaMa-II and a much weaker dependence for PONaMa-I. Figures~\ref{fig:coincidence} and~\ref{fig:AE_sim} compare simulation and data for the two detectors. 

\begin{table}
\begin{center}
\caption{Pulse-shape-discrimination performance of PONaMa-II and PONaMa-II. For this analysis the acceptance of the double-escape peak (mostly single-site events $\rightarrow$ SSE) is fixed to 90\% and the survival fraction of the single-escape peak events (mostly multi-site events $\rightarrow$ MSE) is evaluated. Furthermore we evaluate the fraction of events surviving the cut in full-energy peaks (FEP) that are near the double-escape peak. }
\begin{tabular*}{\textwidth}{@{\extracolsep{\fill}}rrrrr}
\hline
Detector  & MSE (Data)	&  MSE (Sim)	   & FEP (Data)  & FEP	(Sim) \\ 
\hline \hline
PONaMa-I & 4.2  $\pm$ 0.6  & 4.6  $\pm$ 0.2 & 9.3 $\pm$ 0.9 & 12.1 $\pm$ 0.5\\
PONaMa-II & 25.5 $\pm$ 1.4  & 17.9 $\pm$ 0.4 & 35.3 $\pm$ 1.5 & 32.6 $\pm$ 0.8 \\
\hline
\end{tabular*}
\label{tab:lines}
\end{center}
\end{table}
 
\section{Conclusion}
P-type point-contact germanium detectors are ideal for low-background experiments, as they create characteristic pulse-shapes that allow to distinguish signal from background events. This method is heavily exploited by neutrinoless double beta decay experiments such as the \textsc{Majorana Demonstrator}~\cite{Abgrall:2013rze} and the GERDA~\cite{Agostini:2018tnm, Agostini:2013mzu} experiment. To allow for a reliable and efficient background discrimination, a precise understanding of pulse-shapes is inevitable. 

In this work we performed a detailed investigation of pulse-shapes in ultra-high purity P-type point-contact (PPC) Ge-detectors. Low net impurity level are important to deplete large detectors with manageable bias voltages. The studies reveal that too low net impurity levels lead to a degraded pulse-shape discrimination performance. Based on dedicated measurements and accompanying simulations a quantitative description of this behavior could be accomplished. 

Low net impurity levels lead to a reduced electric field, especially in the areas furthest away from the point contact, where the field is dominated by the net impurity concentration. These low field regions lead to slow collection times and hence to a modified pulse-shape for events originating from these areas.

This hypothesis was verified by dedicated measurements with an ultra-low net impurity (between zero and $10^9$~cm$^{-3}$) and a normal-purity ($>10^9$~cm$^{-3}$) PCC detector. A radial scan and a drift-time measurement were performed in order to allow to identify and analyze signals from specific regions in the detector. These investigations confirm that pulse-shapes of single-site events originating from the corners of an ultra-low net impurity detector are significantly distorted. In particular, the background-discriminating parameter A/E is no longer uniform throughout the detector volume. 

These experimental observations were validated by detailed pulse-shape simulations. For the purpose of these investigations a custom simulation code was extended to include the evolution of a finite-size charge-cloud (in contrast to a point-like charge). With this extension a very good agreement between measurement and simulation results could be achieved.

The results presented in this work are instrumental for the data analysis of the ongoing \textsc{Majorana Demonstrator} experiment, and provide important information for the design and understanding of future large-scale PPC Ge-detector-based low-background experiments.

\section*{Acknowledgments}
This work was supported by the U.S. Department of Energy, Office of Science, Office of Nuclear Physics. We gratefully acknowledge helpful discussions and invaluable assistance from the \textsc{Majorana} Collaboration. S.\ Mertens gratefully acknowledges support of a Feodor Lynen fellowship by the Alexander von Humboldt Foundation and support by the Helmholtz Association. A.\ Hegai and C.\ Schmitt would like to thank the DAAD for supporting their stay at Lawrence Berkeley National Laboratory.





\bibliographystyle{elsarticle-num}

\section*{References}
\bibliography{references.bib} 

\end{document}